\documentclass[aps,prd,twocolumn,floatfix,showpacs,a4paper,nofootinbib,amsmath,amssymb]{revtex4}
\usepackage{graphicx} 
\usepackage{dcolumn}  
\usepackage{bm}       
\usepackage{color}
\usepackage{float}    
\usepackage{xspace}          

\newcommand{\be}{\begin{equation}}
\newcommand{\ee}{\end{equation}}
\newcommand{\ba}{\begin{eqnarray}}
\newcommand{\ea}{\end{eqnarray}}
\newcommand\en{\ensuremath{\sqrt{s}}\xspace}
\newcommand\enn{\ensuremath{\sqrt{s_\mathrm{NN}}}\xspace}
\newcommand{\jpsi}{\ensuremath{\mathrm{J}/\psi}\xspace}

\newcommand\pbpb{\ensuremath{\textrm{Pb--Pb}}\xspace}
\newcommand\mean[1]{\ensuremath{\langle#1\rangle}\xspace}
\newcommand\bett{\beta_{\textrm{T}}}
\newcommand\bets{\beta_{\textrm{s}}}
\newcommand\gamt{\gamma_{\textrm{T}}}
\newcommand\mbet{\mean{\bett}}
\newcommand{\RoN}[1]  {\MakeUppercase{\romannumeral #1}}

\newcommand{\lmb}          {\ensuremath{\Lambda}}
\newcommand{\almb}         {\ensuremath{\overline{\Lambda}}}
\newcommand{\Om}{$\Omega^-$}
\newcommand{\Mo}{$\overline{\Omega}^+$}
\newcommand{\X}{$\Xi^-$}
\newcommand{\Ix}{$\overline{\Xi}^+$}

\newcommand\pt{\ensuremath{p_{\textrm{T}}}\xspace}
\newcommand\mpt{\mean{\pt}}
\newcommand\mt{\ensuremath{m_{\textrm{T}}}\xspace}
\newcommand\td{\ensuremath{\textrm{d}}}

\newcommand{\avNcoll}      {\ensuremath{\langle N_\mathrm{coll} \rangle}}
\newcommand{\pizero}          {\ensuremath{\pi^{0}}\xspace}

\newcommand{\pipmns}          {\ensuremath{\pi^{\pm}}}
\newcommand{\pipm}          {\ensuremath{\pi^{\pm}}\xspace}

\newcommand{\kapm}          {\ensuremath{\mathrm{K}^{\pm}}}
\newcommand{\kzero}        {\ensuremath{{\mathrm K}^{0}_\mathrm {S}}}
\newcommand{\p}              {\ensuremath{\mathrm{p}}}
\newcommand{\pbar}         {\ensuremath{\mathrm{\overline{p}}}}
\newcommand{\kstar}        {\ensuremath{{\mathrm K}^{*0}}}
\newcommand{\akstar}      {\ensuremath{\overline{\mathrm K}^{*0}}}
\newcommand{\phm}        {\ensuremath{\phi (1020)}}
\newcommand{\Dpm}          {\ensuremath{\mathrm{D}^{\pm}}\xspace}
\newcommand{\Dzero}        {\ensuremath{{\mathrm D}^{0}}}

\newcommand{\gevc}         {\textrm {GeV}/{\it c}\xspace}

\begin{document}

\title{A three component model for hadron $p_{\rm T}$-spectra \\ in pp and Pb--Pb collisions at the LHC}

\author{Smbat Grigoryan}
\email{Smbat.Grigoryan@cern.ch}
\affiliation{Joint Institute for Nuclear Research, 141980 Dubna, Russia}
\affiliation{A.I.Alikhanyan National Science Laboratory (YerPhI) Foundation, 0036 Yerevan, Armenia}

\date{\today}
\begin{abstract}
A three component model, consisting of the hydrodynamical blast-wave term and two power-law terms, is proposed for an improved fitting of the hadron spectra measured at midrapidity and for arbitrary transverse momenta ($p_\textrm{T}$) in pp and heavy-ion collisions of different centralities at the LHC. The model describes well the available experimental data for all considered particles from pions to charmonia in pp at $\sqrt{s}=$ 2.76, 5.02, 7, 8 and 13~TeV and in Pb--Pb at 
$\sqrt{s_\mathrm{NN}}=$ 2.76 and 5.02~TeV.
\end{abstract}


\maketitle

\section{\label{sec1}Introduction}

Hadron momentum spectra, their dependence on charged-particle multiplicity (equivalently on collision centrality) and their nuclear modification are key observables allowing us to investigate the mechanisms of hadron production in 
proton-proton (pp) and heavy-ion collisions.
It is expected that the heavy-ion collisions produce an initial hot and dense strongly interacting medium (quark-gluon plasma, QGP) which expands (flows) longitudinally and radially, cools and goes through the thermalization, hadronization and finally freezes out to the observable hadrons.
These thermal hydrodynamical processes are responsible mainly for the low and intermediate transverse momentum (\pt) parts of hadron spectra. 
Other processes, such as the QCD hard scatterings and strings fragmentation, production of quark-gluon jets and their energy loss in the medium, 
are responsible mostly for the high-\pt part of spectra. An important contribution to the measured hadron spectra comes from the decays of higher 
mass resonances (see, e.g., Ref.~\cite{Heinz}).\\
\indent
In recent years, different phenomenological models were proposed to describe the hadron momentum spectra. Models based on the power-law Tsallis distribution~\cite{Tsallis} 
describe well the spectra in minimum bias or 
moderate multiplicity pp collisions (see, e.g., Refs.~\cite{Wilk,Biro1,Cley1,Wong,Cley2,Grig,Biro2,Cley3} and references therein). The same is true also for the two component model based on the sum of Boltzmann-Gibbs distribution and a power-law term ~\cite{Byl1}. But for heavy-ion collisions and high multiplicity pp collisions these models fail to describe the measured spectra in the whole available 
\pt range and more complex models were suggested~\cite{Byl2,Biro3,Zhu1,Eloss,Liu,Gupta}, 
some of which include the hydrodynamical flow effects via the blast-wave model (BWM)~\cite{BW}. 
The BWM alone is often used to fit the spectra in a limited \pt range~\cite{Cley3,ali1,Melo,Maz1}(e.g., \pt $\sim$ 0.5--3~\gevc~\cite{Maz1}).\\
\indent
In the present paper, a three component model is proposed to fit accurately the hadron \pt-spectra at midrapidity and for any \pt measured in heavy-ion and pp collisions at the LHC. The model (further referred to as BWTPM) consists of a standard BWM term, a Tsallis term and another power-law term (see Eq. (2) below). 
Unlike to the similar three component model~\cite{Byl2}, in our model the two power-law terms have different forms and 
parameters $\bets$ and $k$ of the BWM term are fitting parameters, while they are equal to 0.5 and 1, respectively, in Ref.~\cite{Byl2}. 
These differences allow a better description of the LHC data. Considered data include
the \pt-spectra of \pipm, \pizero, \kapm, \kzero, \p, \pbar,  \kstar, \akstar, $\phi$, \lmb, \almb, \X, \Ix, \Om, \Mo, \Dzero, \Dpm, \jpsi 
and unidentified charged-particles in pp and Pb--Pb collisions at different centralities and  
\en (in this paper \en denotes both the pp collision energy and the Pb--Pb collision energy per nucleon-nucleon pair, though the latter also denoted as \enn). By examining the data fits it is observed that the model parameters depend on the collision system (pp or Pb--Pb), 
but some are independent of the \en or
the particle type or the collision centrality. Therefore, to
determine the parameters, only two global fits are 
performed: one for the pp data and another one for the Pb--Pb data. Other collision systems will be considered elsewhere.

\section{\label{sec2}Model description}

For a particle \pt-spectrum at midrapidity, where it can be considered as independent of the rapidity ($y$), 
we propose the following model

\be \label{eq:d2N}
\frac{\td^2 N}{\td\pt\td y} \equiv  Y(\pt) = \pt \left\{
\begin{array}{ll}
F(\pt) \quad  \pt \leq p_h,\\
h(\pt) \quad  \,\pt > p_h,
\end{array}
\right. 
\ee

\vspace{-1.2em}

\begin{eqnarray} \label{eq:F}
&& \hskip -1.0em F(\pt) = \frac{g}{\pi^2}\mt \left[f_{BW} V \int_0^1 x dx I_0(\gamt \bett \frac{\pt}{T}) K_1(\gamt\frac{\mt}{T})\right.  \nonumber\\
&& \hskip 1.0em + \left.f_1 V_1(1+ \frac{\mt}{c_1 e_1 n})^{-c_1 n} + f_2 V_2(1 +\frac{\pt^2}{c_2^2 e_2^2})^{-\frac{n}{2}}\right], 
\end{eqnarray}
where N is the particle yield per collision event, $g=2J+1$ is its spin degeneracy factor,  
$\mt=\sqrt{m^2+\pt^2}$ is its transverse mass, $\gamt=1/\sqrt{1-\bett^2}, \bett = \bets x^k$ is the medium radial (transverse) flow velocity with the profile exponent $k$ and surface velocity $\bets$, $x=r/R, R$ is the upper boundary of  
radial coordinate $r$, $V=\pi R^2\tau_f$, $\tau_f$ and $T$ are the medium proper time and temperature at kinetic freeze-out, $I_0$ and $K_1$ are the modified Bessel functions.
The data fits show that parameters 
$f_{BW}, f_1, f_2, c_1 \geq 1, e_1$ and $c_2$ depend
only on the particle type and collision system. 
Parameters $V, V_1, V_2, T, k, e_2$ and $n$ depend only on the collision system, centrality and \en, while $\bets$ depends only on the collision system and \en.

The model main function $F(\pt)$ has three terms. 
First term, representing the thermal part of the spectrum in the standard BWM form~\cite{BW}, decreases exponentially at high \pt and is important at low and intermediate values of \pt. Second therm in the Tsallis distribution form~\cite{Tsallis} is 
more significant for pions and kaons at low \pt and presumably is responsible mostly for the low-\pt contribution
from the heavier resonance decays (the rest of the decay contribution is included in all terms of Eq.~(\ref{eq:F})). 
Third therm in the $\pt^2$-dependent power-law form
(used, e.g., in Refs.~\cite{Wong,Byl1,Byl2,Grig5}) describes the QCD hard processes and is most important in the
high-\pt region. The characteristic energy scales for these three terms are
$T \sim$ 133--162~MeV, $e_1 \sim$ 42--695~MeV 
and $c_2 e_2 \sim$ 0.7--4.2~GeV, respectively
(see Sec.~\ref{sec3} for the parameter values).\\
\indent
Function $F(\pt)$ fits with a high accuracy almost all the available pp and Pb--Pb LHC data in the region $\pt \leq p_h = 40$~\gevc and has a \mbox{high-\pt} behavior $\propto {\pt}^{1-n}$, where the power index $n$ depends on the collision system, centrality and \en. However, it is expected from the QCD~\cite{Wong,Arleo} and confirmed by the data (see the good fits at $\pt > p_h$ in the Sec.~\ref{sec4}) that hadronic spectra should have an universal high-\pt behavior $\propto {\pt}^{1-n_h}$ with $n_h$ depending only on \en. To provide such a behavior and to continue $F(\pt)$ smoothly into the region $\pt > p_h$, the following simple form for the function $h(\pt)$ in Eq.~(\ref{eq:d2N}) is chosen
\be  \label{eq:hpt}
h(\pt)=F(p_h)[1+\alpha(1-x)]\,x^{n_h-1+(n-n_h)x^3},
\ee
where $x=p_h/\pt$ and parameter $\alpha$ depends on centrality in case of Pb--Pb collisions. 
Function $h(\pt)$ with $\alpha = 0.11(\mathrm{GeV}/e_2)^5$ and values of $e_2$, $n$ and $n_h$ given in the Sec.~\ref{sec3}, fits well the available scarce data at $\pt > p_h$ (including the spectra of charged-particles and few other hadrons). Due to the Eq.~(\ref{eq:hpt}), the ratio of any two \pt-spectra of same \en will reach a plateau at high \pt.

The unidentified charged-particle spectrum is a sum of the $\pipm, \kapm, \p, \pbar$ and higher mass charged hyperons spectra and is defined, assuming same spectrum for a particle and its anti-particle at the LHC energies, by the following equation
\be  \label{eq:chp}
\frac{\td^2 N_{ch}}{\td\pt\td\eta} = 2(\sum\limits_{i=\pi,\mathrm{K,p}}^{} \frac{\pt}{{\mt}_{,i}}Y_i + f_{hyp}\frac{\pt}{{\mt}_{,\mathrm p}}Y_\mathrm{p})\, .
\ee
Here, the factor 2 accounts for the positive and negative particles,  the factors $\pt/\mt$ account for the change from the rapidity to pseudo-rapidity ($\eta$) at midrapidity ($d\eta/dy \approx \pt/\mt$)
and the last term describes approximately the small contribution of hyperons via the proton
contribution, scaled by the factor $f_{hyp} =$ 0.3384 (0.3772) determined from the fits for pp (Pb--Pb) collisions. 

\section{\label{sec3}Fitted data and parameters}

Two global fits of different hadrons \pt-spectra at midrapidity are performed using Eqs.~(\ref{eq:d2N})--(\ref{eq:chp}): one for the pp data~\cite{ali2,ali3,ali4,ali5,ali6,ali7,ali8,ali9,ali10,ali11,cms1,ali12,ali13,ali14,atl1,atl2,cms2,cms3,cms4,ali15,ali16,ali17,ali18,ali19,ali20,ali21,ali22,atl3,cms5,cms6}
and another one for the Pb--Pb data~\cite{ali18,ali19,ali20,ali21,ali22,atl3,cms5,cms6,ali23,ali24,ali25,ali26,ali27,cms7,atl4}, 
measured by the ALICE (mostly), \mbox{ATLAS} and CMS experiments at the LHC. 
By midrapidity we generally mean $|y|<1$, 
but we included in the fit also the charged-particle spectrum at \en $=$ 7~TeV and $|\eta|<2.4$~\cite{cms1}, which can be considered as midrapidity for not too high \pt.
The fitted pp data include the charged-particle multiplicity dependent measurements at \en $=$ 7 and 13~TeV for ten
so called V0M multiplicity classes in INEL$>$0 events (having at least one charged-particle in $|\eta|<1$), defined in Refs.~\cite{ali2} and~\cite{ali3}, respectively.
Also the data of minimum bias inelastic (INEL) pp collisions at \en $=$ 2.76, 5.02, 7, 8 and 13~TeV are included.
The hadron spectra, measured as cross sections, are transformed to hadron yields per event using the
inelastic cross section

\begin{table}[h]
   \caption{The collision energy and centrality depending fitting parameters for ten V0M multiplicity classes in INEL$>$0 pp collisions at \en $=$ 7 and 13~TeV, defined in Refs.~\cite{ali2} and~\cite{ali3}, respectively. 
These parameters are given also for pp inelastic events (INEL) at five values of \en. Values of $n$ should be multiplied by 1.09 for \jpsi.} \label{tab1}
  \begin{tabular}{*{7}{c}}
    \colrule
    \en  & Class & $V ({\rm fm}^3)$ & $V_1 ({\rm fm}^3)$ & $V_2 ({\rm fm}^3)$ & $k$ &  $n$ \\
    \colrule
2.76 TeV   & INEL & 19.02 & 258.2 & 0.005363 & 2.768 & 7.636 \\
   \colrule
5.02 TeV   & INEL & 20.55 & 275.8 & 0.006463 & 2.606 & 7.335 \\
    \colrule
8 TeV   & INEL & 21.68 & 303.6 & 0.006992 & 2.441 & 7.192 \\
    \colrule
     & I & 64.51 & 1475 & 0.04334 & 0.5674 & 6.891 \\ 
     & II & 55.87 & 1157 & 0.03341 & 0.8572 & 7.031 \\
     & III & 49.33 & 941.5 & 0.02676 & 1.170 & 7.136 \\
    & \RoN{4} &  45.55 & 795.0 & 0.02239 & 1.485 & 7.208 \\  
  7 TeV & V & 42.28 & 689.1 & 0.01870 & 1.785 & 7.245 \\
    & VI & 37.97 & 572.0 & 0.01509 & 2.265 & 7.314  \\
    & \textrm{VII} & 32.90 & 446.5 & 0.01084 & 2.989 & 7.371  \\  
    & VIII & 27.90 & 358.1 & 0.007537 & 3.764 & 7.396 \\
    & IX & 21.65 & 260.2 & 0.003759 & 5.010 & 7.365 \\
    & X & 13.51 & 153.1 & 0.000654 & 10.52 & 7.170 \\
    & INEL & 21.33 & 302.4 & 0.006688 & 2.497 & 7.213 \\
    \colrule
     & I & 75.49 & 1594 & 0.05427 & 0.4669 & 6.664 \\
     & II &  64.56 & 1287 & 0.04285 & 0.7393 & 6.813 \\
     & III & 57.09 & 1039 & 0.03474 & 1.060 & 6.919 \\
    & IV & 52.78 & 872.0 & 0.02851 & 1.373 & 6.982 \\
  13 TeV & V & 48.61 & 756.1 & 0.02470 & 1.689 & 7.043 \\
    & VI & 42.60 & 635.2 & 0.02070 & 2.203 & 7.144 \\
    & VII & 37.15 & 490.6 & 0.01546 & 2.977 & 7.240 \\
    & VIII &  31.10 & 391.8 & 0.01152 & 3.849 & 7.345 \\
    & IX & 23.61 & 280.2 & 0.006509 & 5.227 & 7.426 \\
    & X & 14.34 & 152.9 & 0.001322 & 11.71 & 7.324 \\
    & INEL & 22.35 & 307.1 & 0.009059 & 2.261 & 7.086 \\
    \colrule
   \end{tabular}
\end{table}

\begin{widetext}

\begin{table*}[h]
   \caption{The collision energy and centrality depending model parameters for different centrality classes in \pbpb collisions at \enn $=$ 2.76 and 5.02 TeV defined 
   in Ref.~\cite{ali28}. Values of $n$ should be multiplied by 1.09 for \jpsi.} \label{tab2}
  \begin{tabular}{*{9}{c}}
    \colrule
    \enn  & Class & $V ({\rm fm}^3)$ & $V_1 ({\rm fm}^3)$ & $V_2({\rm fm}^3)$ & $k$ & $T$ (GeV)&  $n$ & $e_2$ (GeV) \\
    \colrule
     & 0--5\% & 6617 & 36024 & 168.3 & 0.1696 & 0.1380 & 6.801 & 0.6973 \\
     & 5--10\% & 4770 & 29957 & 141.4 & 0.1837 & 0.1400 & 6.902 & 0.7307 \\
     & 10--20\% & 3125 & 23028 & 104.6 & 0.2214 & 0.1425 & 6.975 & 0.7665 \\
    & 20--30\% & 1874 & 15902 & 69.49 & 0.2699 & 0.1444 & 7.001 & 0.7901 \\
  2.76 TeV & 30--40\% & 1203 & 10718 & 36.91 & 0.4058 & 0.1462 & 7.053 & 0.8371 \\
    & 40--50\% & 816.8 & 6806 & 15.90 & 0.6326 & 0.1465 & 7.108 & 0.9083 \\
    & 40--60\% & 704.9 & 5345 & 10.58 & 0.7343 & 0.1452 & 7.130 & 0.9386 \\
    & 60--80\% & 312.5 & 1366 & 1.038 & 1.346 & 0.1397 & 7.334 & 1.154 \\
    \colrule
	& 0--5\% & 8201 & 35193 & 317.3 & 0.0790 & 0.1334 & 6.580 & 0.6505 \\
	& 5--10\% & 5892 & 29109 & 253.0 & 0.0899 & 0.1355 & 6.624 & 0.6722 \\
	& 10--20\% & 3689 & 23288 & 178.2 & 0.1173 & 0.1388 & 6.714 & 0.7147 \\
	& 20--30\% & 2027 & 16918 & 117.2 & 0.1717 & 0.1432 & 6.752 & 0.7390 \\
  5.02 TeV  & 30--40\% & 1160 & 11535 & 71.88 & 0.2826 & 0.1469 & 6.769 & 0.7643 \\
	& 40--50\% & 694.2 & 7660 & 36.45 & 0.4901 & 0.1497 & 6.809 & 0.8076 \\
	& 50--60\% &  436.4 & 4700 & 15.28 & 0.8374 & 0.1510 & 6.857 & 0.8657 \\
	& 60--70\% &  282.9 & 2517 & 4.602 & 1.305 & 0.1498 & 6.959 & 0.9748 \\
	& 70--80\% & 167.2 & 1170 & 1.205 & 1.807 & 0.1475 & 7.058 & 1.089 \\
	& 80--90\% & 57.38 & 465.8 & 0.4796 & 2.596 & 0.1511 & 7.093 & 1.078 \\
    \colrule
   \end{tabular}
\end{table*}

\begin{table*}[h]
   \caption{The particle type and collision system depending model parameters. } \label{tab3}
  \begin{tabular}{*{12}{c}}
    \colrule
   System  & Parameter \; & \pipmns (\pizero)\; & \;\kapm, \kzero & \p, \pbar &  \kstar, \akstar & \phm &\lmb, \almb  & \X, \Ix & \Om, \Mo & \Dzero, \Dpm & \jpsi \\
    \colrule
     & $f_{BW}$ & 1 & 0.3539  & 0.7653  &  0.4186 & 0.1970  & 1.192  & 0.2833  & 0.0570  &  8.325 & 15.01 \\
     & $f_1$      & 1(1.30) & 0.02205& 0.005113  & 0  &  $2.603\!\cdot\!\!10^{-5}$  &  0 &  0 & 0  & 0  & 0  \\
   pp  &  $f_2$  & 1(1.09) & 0.4964  & 0.2353  & 0.04705  & 0.01447  & 0.2885  &  0.02641 & 0.0009601 &  0.007366 & $2.241\!\cdot\!\!10^{-5}$ \\
     &  $c_1$   & 1.000  & 1.401 &  $\infty$ & -  &  $\infty$ & -  & -  & -  &  -  & -  \\    
    & $e_1$ (GeV) &  0.04156 & 0.09924  &  0.1627 &  - & 0.3115  & -  &  - & -  &  - & -  \\
    &  $c_2$        & 1 &  1.011 & 0.8476   & 1.173  &  1.281 & 0.8129  & 0.9702  & 1.121  &  1.850 & 2.320  \\
     \colrule
     & $f_{BW}$ & 1 & 1.000  & 2.690  & 0.8681  & 0.8662  & 5.622  &  2.335 &  1.299 &  112.5  & 500.0 \\
     & $f_1$       & 1 & 0.01294  &  $1.511\!\cdot\!\!10^{-6}$ &  0 & 0  &  0 & 0 & 0  &  0 & 0  \\
   \pbpb &  $f_2$ & 1 & 0.09365  & 0.005177  & 0.001405  & 0.001592  & 0.003920  & 0.0007995  &  $4.817\!\cdot\!\!10^{-5}$ &   0.0006266 & $2.050\!\cdot\!\!10^{-6}$ \\
    &  $c_1$        &  1.488 & $\infty$  & $\infty$  &  - &  - &  - & -   & -   & -  & -   \\
    & $e_1$ (GeV) & 0.06280  & 0.1581  & 0.6950  & -  &  - & - &  - & -  & - & -    \\
    &  $c_2$        & 1 &1.292   &  1.374 &  2.036 &  1.777 & 1.561  &  1.571 &  1.858 &  2.901 & 3.600  \\
    \colrule
   \end{tabular}
\end{table*}

\end{widetext}

\begin{table}[ht]
   \setlength\tabcolsep{-1.0pt} 
   \caption{$\bets$ and $n_h$ versus \en  and collision system.} \label{tab4}
 \begin{tabular*}{0.49\textwidth}{@{\extracolsep{\fill}}*{8}{c}}
    \colrule
    System & \multicolumn{5}{c}{pp} & \multicolumn{2}{c}{Pb--Pb}  \\ 
       \cline{1-1} \cline{2-6} \cline{7-8}
    \en (TeV) &\, 2.76 & 5.02 & 7 & 8 & 13 & 2.76 & 5.02\\
       \cline{1-1} \cline{2-6} \cline{7-8}
   $\bets$  &\, 0.7330 & 0.7577 & 0.7707 & 0.7758 & 0.7937 & 0.7698 & 0.7924 \\
       \cline{1-1} \cline{2-6} \cline{7-8}
   $n_h$  &\, 7.94 & 7.63 & 7.50 & 7.47 & 7.36 & 7.94 & 7.63 \\
    \colrule
   \end{tabular*}
\end{table}

\noindent
values from Ref.~\cite{Loiz} for the LHC energies.\\
\indent
The used Pb--Pb data include measurements  at \enn $=$ 2.76 and 5.02 TeV for different centrality classes (see Table II), corresponding to different multiplicities of charged-particles and defined as the percentiles of the Pb--Pb hadronic cross section~\cite{ali28}.
The charged-particle and other hadron multiplicities decrease when going from the most central class (0--5\%) to peripheral ones. Note that to fit the data for a large centrality interval, being a sum of several centrality
classes, we use the average of the fit functions of these classes. For instance, data for the 0--10\% centrality can be fitted with the arithmetic average of two fit functions of classes 0--5\% and 5--10\%.\\
\indent
The resulting ratios $\chi^2/NDF$ for the pp and Pb--Pb global fits  (in the ROOT framework~\cite{ROOT}) are 894.7/6530 and 866.0/5160, respectively. 
Values for the most of fitting parameters are given in Tables I--IV. Parameters $T$ and $e_2$ for pp collisions, unlike for Pb--Pb collisions, show independence of the collision energy and centrality and are defined from the fit as: $T=0.1618$~GeV, $e_2 = 1.7375$~GeV.
Parameters $f_{BW}, f_1, f_2$ and $c_2$ for \pipm are fixed equal to unity. The \pizero and \pipm have the same parameters except the $f_1$ and $f_2$ for pp collisions. Their values for \pizero are given in parentheses in the Table III. To observe such difference in Pb--Pb collisions one needs more \pizero data.
For cases with $c_1 = \infty$ the Tsallis form in the second term of Eq.~(\ref{eq:F}) should be replaced by an exponential due to $\lim_{a\to\infty} (1+x/a)^a = e^x$.
Since \Dpm yields are lower than for \Dzero, \Dpm have, in addition to the parameters given in 
the Table III, a normalization parameter equal to 0.437 (0.413) for pp (Pb--Pb) collisions. 
Fit of the CMS charged-particle \pt-spectrum for non-single-diffractive pp collisions at \en $=$ 7~TeV~\cite{cms1}
needs an additional normalization parameter of 1.16 with respect to the corresponding ALICE INEL data~\cite{ali11}.
In fact, besides the parameters of Table III, other parameters are common for all particles.
Only for \jpsi the values of parameter $n$ in Tables I and II should be multiplied by 1.09.
Note that in the fits we use the spectra of prompt $\mathrm{D}$ and \jpsi mesons, not including contributions from the decays of heavier hadrons, containing $b$-quarks.
The large values of $f_{BW}$ for these mesons in the Table III remind us the charm quark large fugacity obtained in the 
statistical hadronization model~\cite{shm1,shm2}.

It is worth to mention that many of the papers, which utilize BWM for the centrality dependent data fits, are using $\mbet$ instead of $\bets$ as a fitting parameter,
where $\mbet = 2\bets/(k+2)$ is the the radial flow velocity averaged over the radial coordinate. Since our fits give centrality independent $\bets$, 
the centrality dependence of $\mbet$ is defined by parameter k. As a result, $\mbet$ grows strongly (weakly) with increasing collision centrality (energy).\\
\indent
Let us discuss the individual contributions of three components of Eq.~(\ref{eq:F}) into the different particles \pt-spectra. Figure 1 presents
the first, second, third components and the total spectrum, denoted as BW, Ts, Po and Tot, respectively, for most central (full lines) and most peripheral (dashed lines) collisions in pp at \en$=$7~TeV and in Pb--Pb at \enn$=$5.02~TeV. Notice the following properties of three components:

\onecolumngrid  

\begin{figure}[h]
\includegraphics[width=1.0\columnwidth]{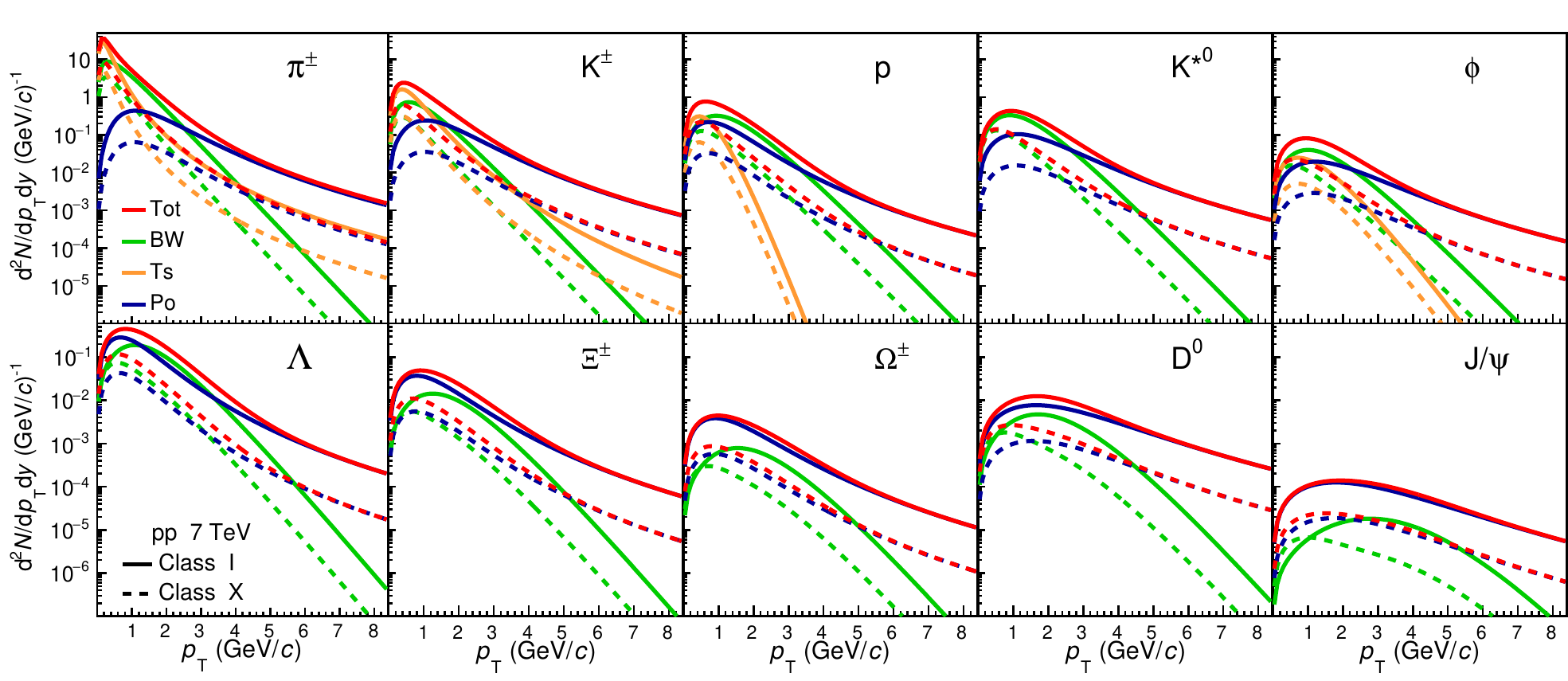}
\vskip -0.5mm
\includegraphics[width=1.0\columnwidth]{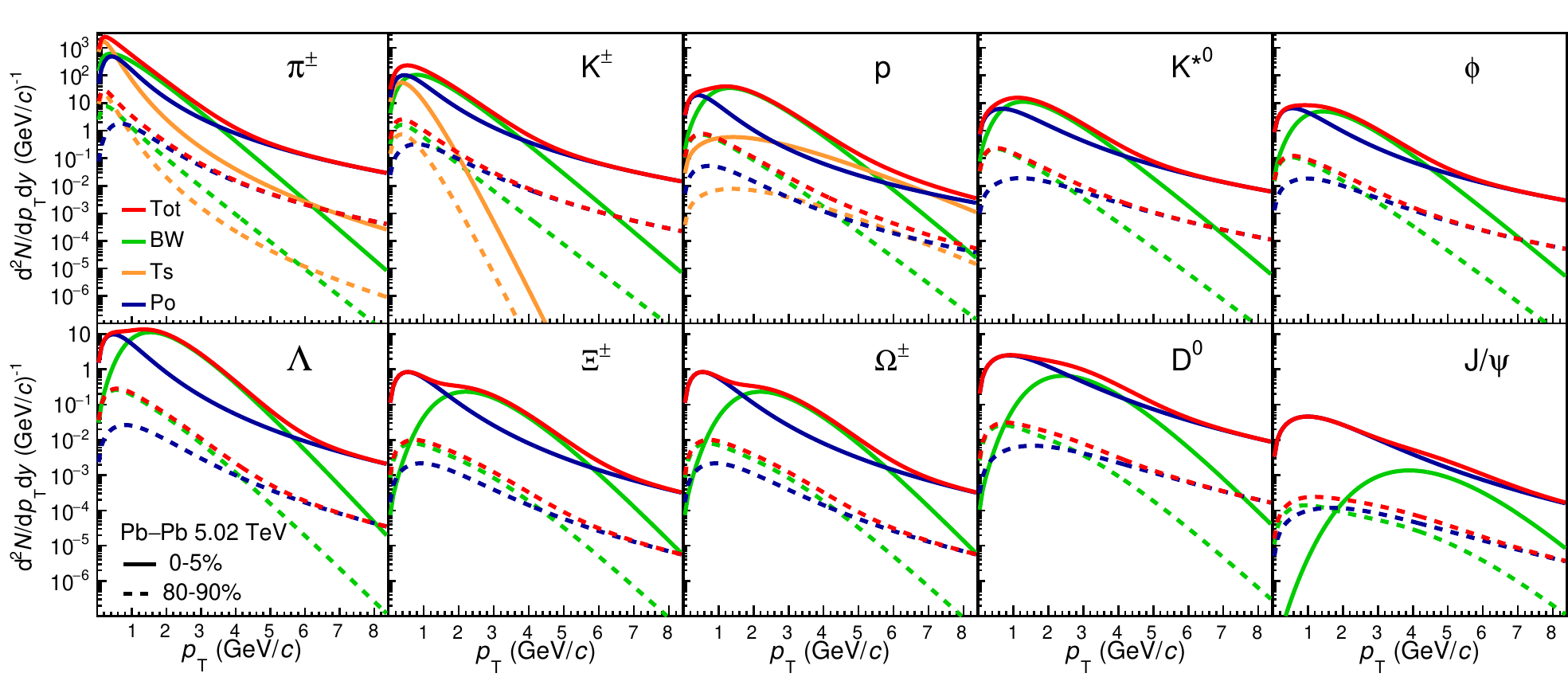}
\vskip -2mm
\caption{Three components of Eq.~(\ref{eq:F}) and their sum (denoted as BW, Ts, Po and Tot, see the text) for different particles in case of most central (full lines) and most peripheral (dashed lines) collisions in pp at \en$=$7~TeV (top panels) and in Pb--Pb at \enn$=$5.02~TeV (bottom panels).} \label{compons}
\end{figure}

\twocolumngrid  

\noindent
1) First one (BW) is important in the low and intermediate
\pt regions, where it has a peak. In the low-\pt region BW is more significant for pp and peripheral Pb--Pb
collisions. The peak gradually shifts from lower to higher \pt values with increasing particle mass or collision centrality.
This is due to the collective radial flow which pushes the particles and increases their momenta 
according to the relation (non-relativistic case) $\mpt \approx \mpt_{0} + m\mbet$~\cite{Heinz,BW}, 
where $ \mpt_{0}$ is the mean transverse momentum without the flow, $\mbet$ is larger for Pb--Pb than for pp and grows with the collision centrality.\\
2) Second one (Ts) dominates for pions and kaons at very low \pt. For pions in pp collisions it is significant at any \pt.
The importance of Ts for pions can be explained by the fact that they have the largest contribution
from the resonance decays.
Since the \pt distributions of the secondary and primary pions can differ strongly, both could not be described with only BW and Po components.
For heavier particles the Ts contributions decrease with increasing mass and become negligible. This can be
explained by the smaller feedback (if any) from the resonances into the measured \pt-spectra and by the similarity of the primary and secondary particles spectra.
The larger Ts for protons in Pb--Pb collisions is related possibly to some specific mechanisms of the proton production in these collisions~\cite{shm3}.\\
3) Third one (Po) generally dominates in high-\pt region, which is the domain of the power-law QCD processes. But it is significant also at low \pt, especially for the central Pb--Pb collisions, which is mostly due to the centrality dependence of the energy parameter $e_2$.
This softening of the \pt-spectra with increasing centrality can be explained by the energy loss of quark-gluon jets in the QGP.
In case of \jpsi, its formation via the charm quark-antiquark recombination in the QGP~\cite{shm1} could also be important at low \pt.
For \jpsi the Po component dominates almost always.\\

\section{\label{sec4}Discussion of the results}

\begin{figure}[ht]
\includegraphics[width=1.1\columnwidth]{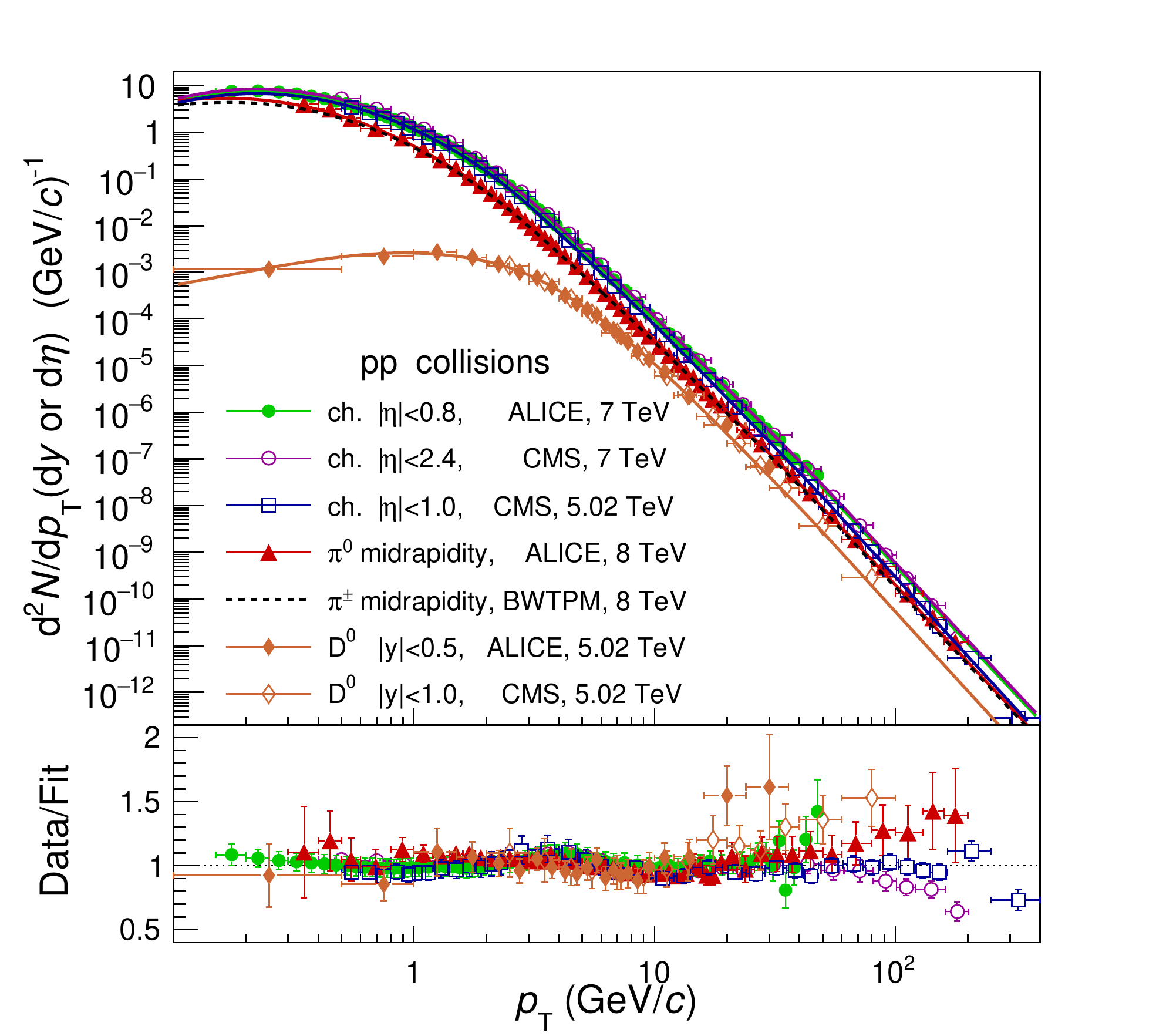}
\vskip -2mm
\caption{Fit of \pt-spectra with highest \pt reach in pp collisions for charged-particles at \en$=$7~TeV~\cite{ali11,cms1} and 5.02~TeV~\cite{cms5}, for \pizero at \en$=$8~TeV~\cite{ali17} and for \Dzero\, at \en$=$5.02~TeV~\cite{ali13,cms6}. 
Model prediction for \pipm at \en$=$8~TeV (dashed line) is shown to compare with \pizero. Ratios data/fit here and in the following Figures demonstrate the quality of the fits.}
\label{ch7pi8}
\end{figure}

Here, several plots are presented to illustrate the results and quality of the fits. Figures 2 and 3 show the fits of different data sets with highest \pt reach for pp and Pb--Pb collisions, respectively.
To demonstrate the quality of the fits, the data points have been divided by the corresponding fit function values, and these ratios are also plotted in the bottom panels. Generally, the quality is always good within the data uncertainties. In particular, the good fits of the charged-particle spectra in both pp and Pb--Pb collisions at 
\en$=$5.02~TeV and $\pt < 400$~\gevc confirm the need of a universal high-\pt behavior in Eq.~\ref{eq:hpt}. Note that the CMS charged-particle spectrum for $|\eta|<2.4$ and \en$=$7~TeV~\cite{cms1} is systematically lower than the model curve at $\pt > 80$~GeV/$c$. 
This could mean that for such kinematics our "midrapidity model" fails and one should take into account the (pseudo)rapidity dependence of \pt-spectra. Figure~\ref{ch7pi8} includes also our prediction for \pipm at 
\en$=$8~TeV. A small

\begin{figure}[H]
\includegraphics[width=1.1\columnwidth]{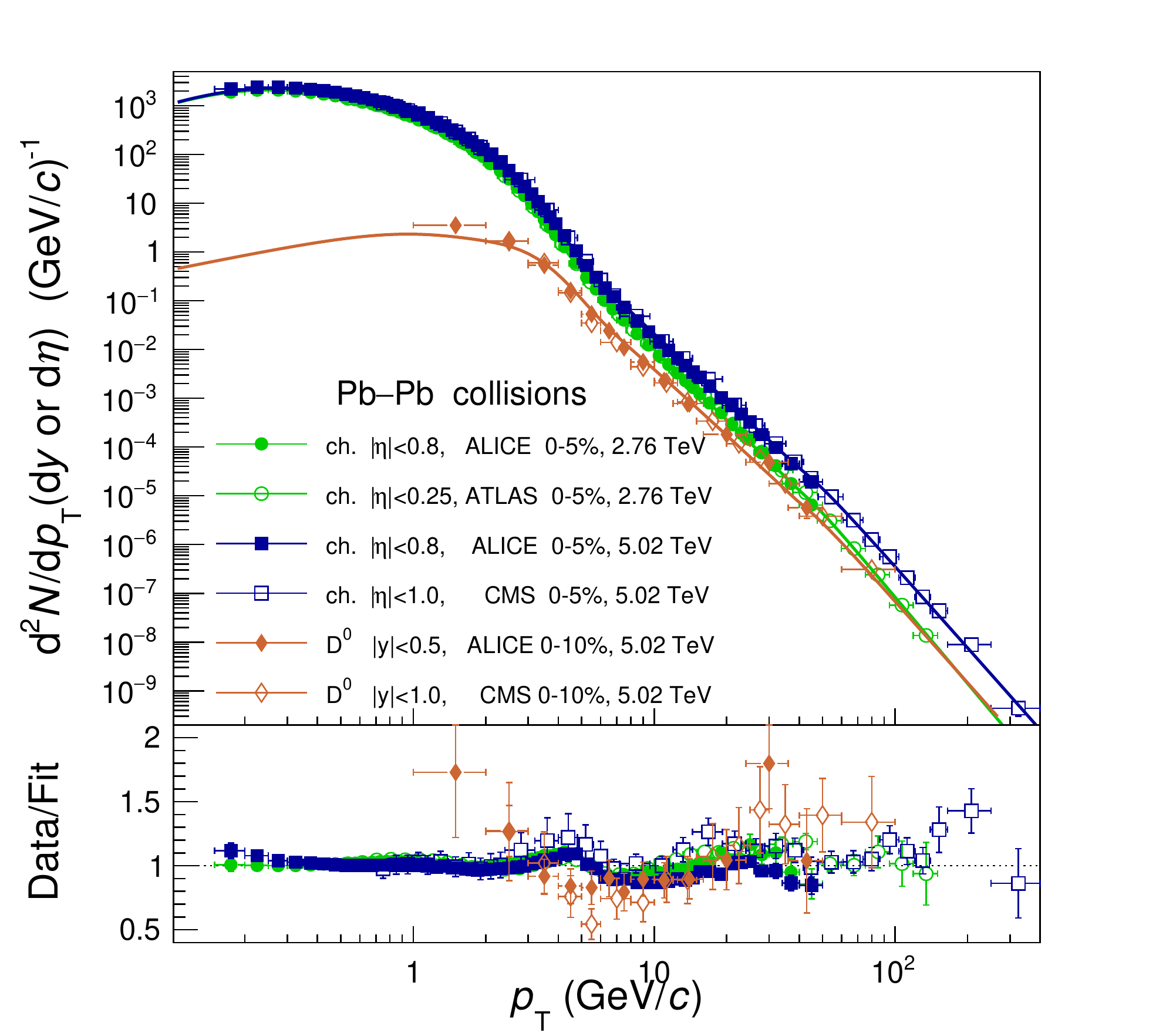}
\vskip -2mm
\caption{Fit of \pt-spectra with highest \pt reach in Pb--Pb collisions for charged-particles at \enn$=$2.76~TeV~\cite{ali20,atl3} and 5.02~TeV~\cite{ali20,cms5} and for \Dzero\, at \enn$=$5.02~TeV~\cite{cms6,ali27}.}
\label{chD0}
\end{figure}

\noindent
difference between \pizero and \pipm yields is probably due to the difference of contributions from the resonance decays.
Similar differences should be also at other LHC energies.\\
\indent
Figures 4 and 5 demonstrate our main fits of the centrality 
dependent spectra for most abundantly produced pions, kaons
and protons in the pp and Pb--Pb collisions.

\onecolumngrid  

\begin{figure}[H]
\includegraphics[width=1.0\columnwidth]{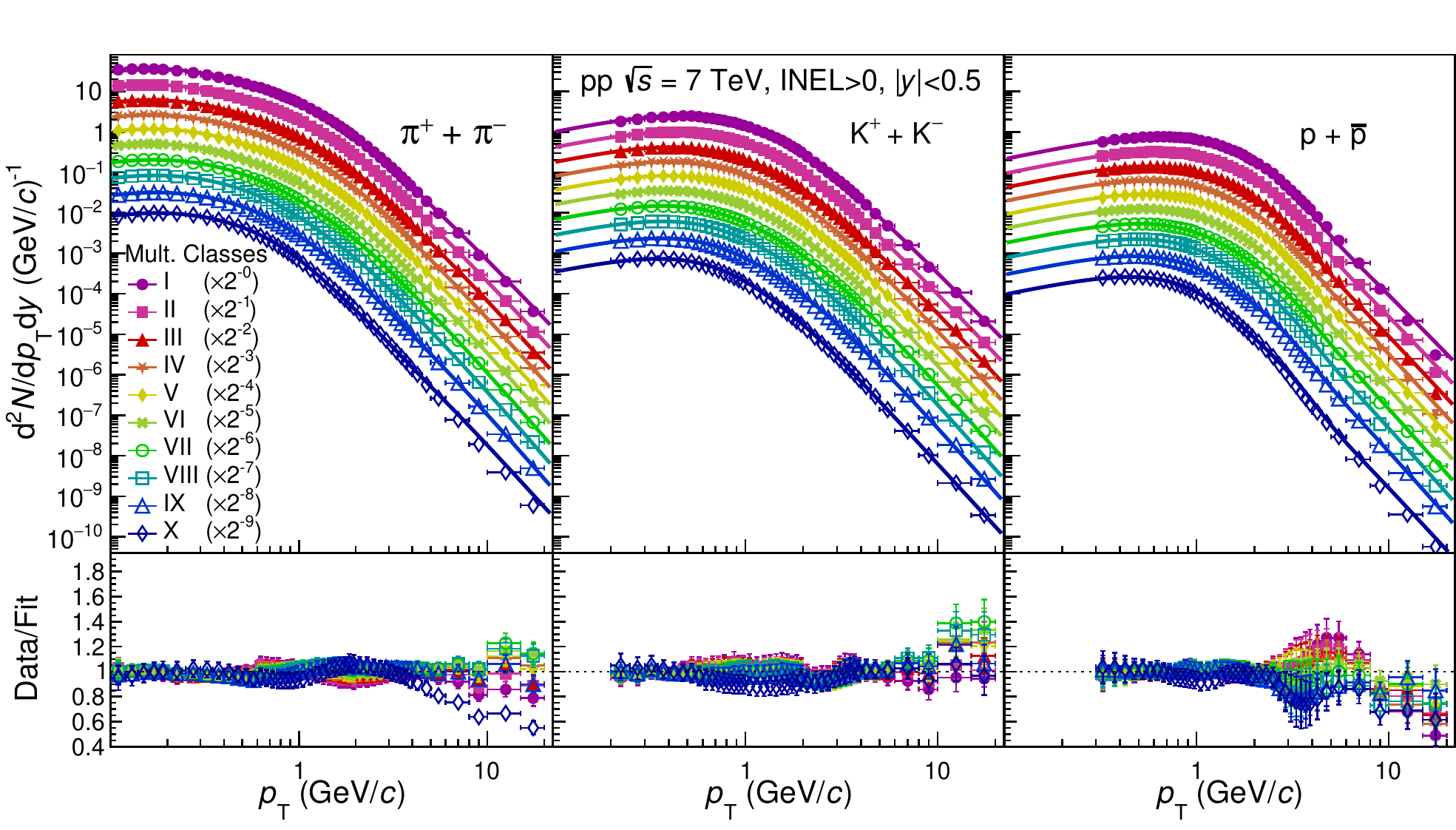}
\vskip -0.5mm
\includegraphics[width=1.0\columnwidth]{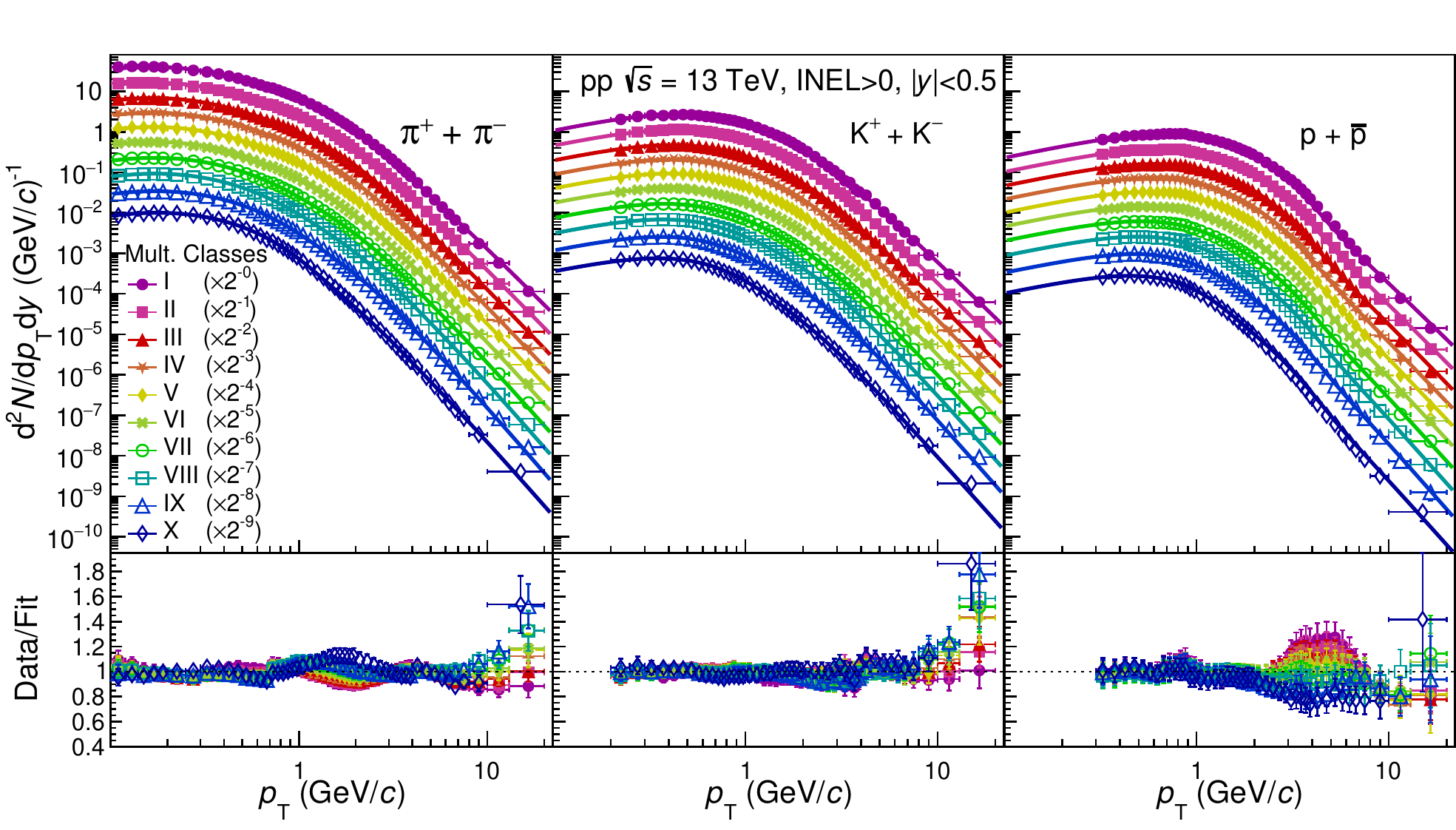}
\vskip -2mm
\caption{Fit of pion, kaon and proton \pt-spectra at $|y| < 0.5$ for ten multiplicity classes of INEL$>$0 events in pp collisions at \en$=$7~TeV~\cite{ali2} and \en$=$13~TeV~\cite{ali3}. 
The data points and the fitting curves in the upper panels are scaled by the numbers given in the parentheses for a better visibility.}
\label{idpvsM}
\end{figure}

\begin{figure}[H]
\includegraphics[width=1.0\columnwidth]{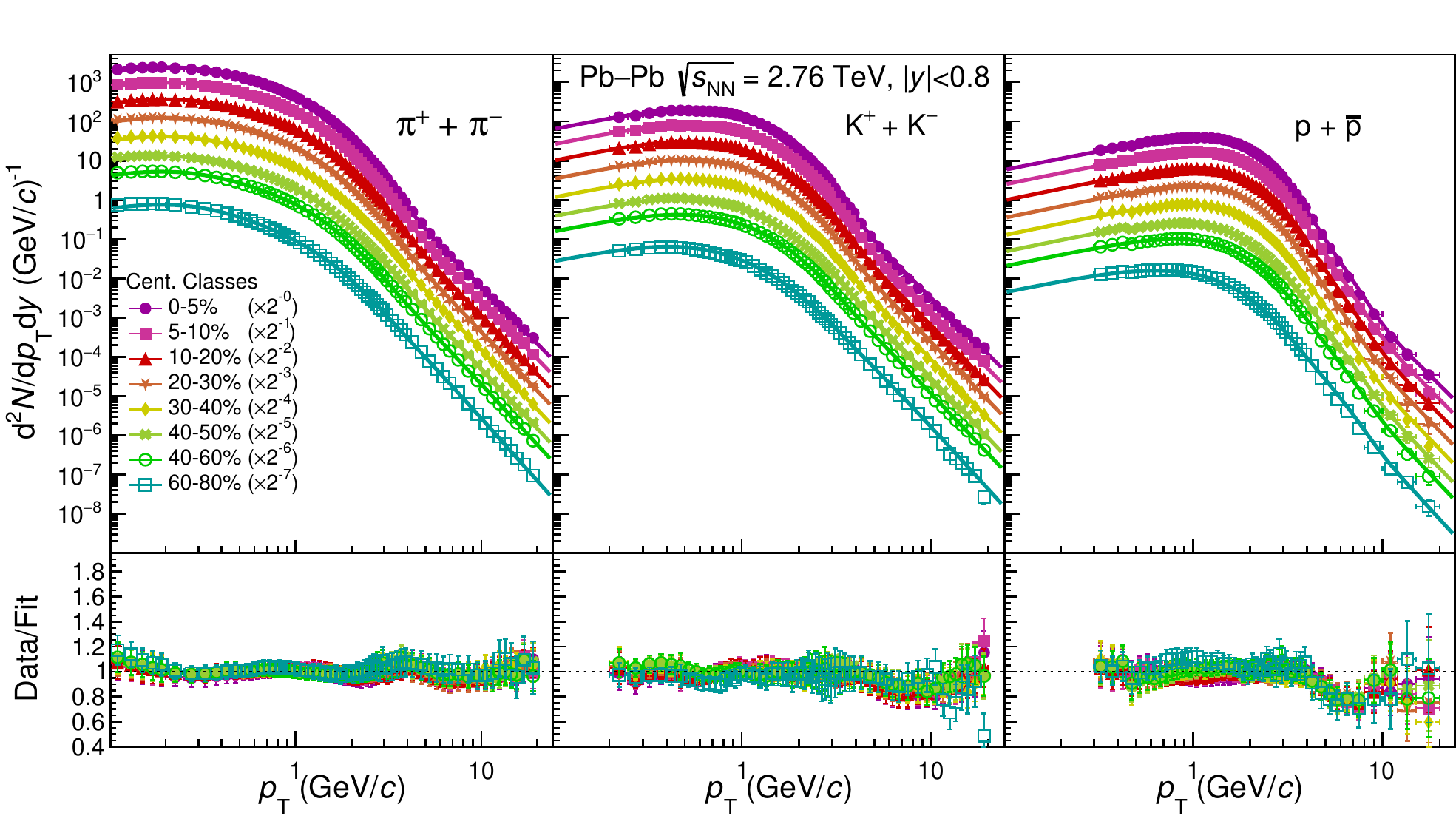}
\vskip -0.5mm
\includegraphics[width=1.0\columnwidth]{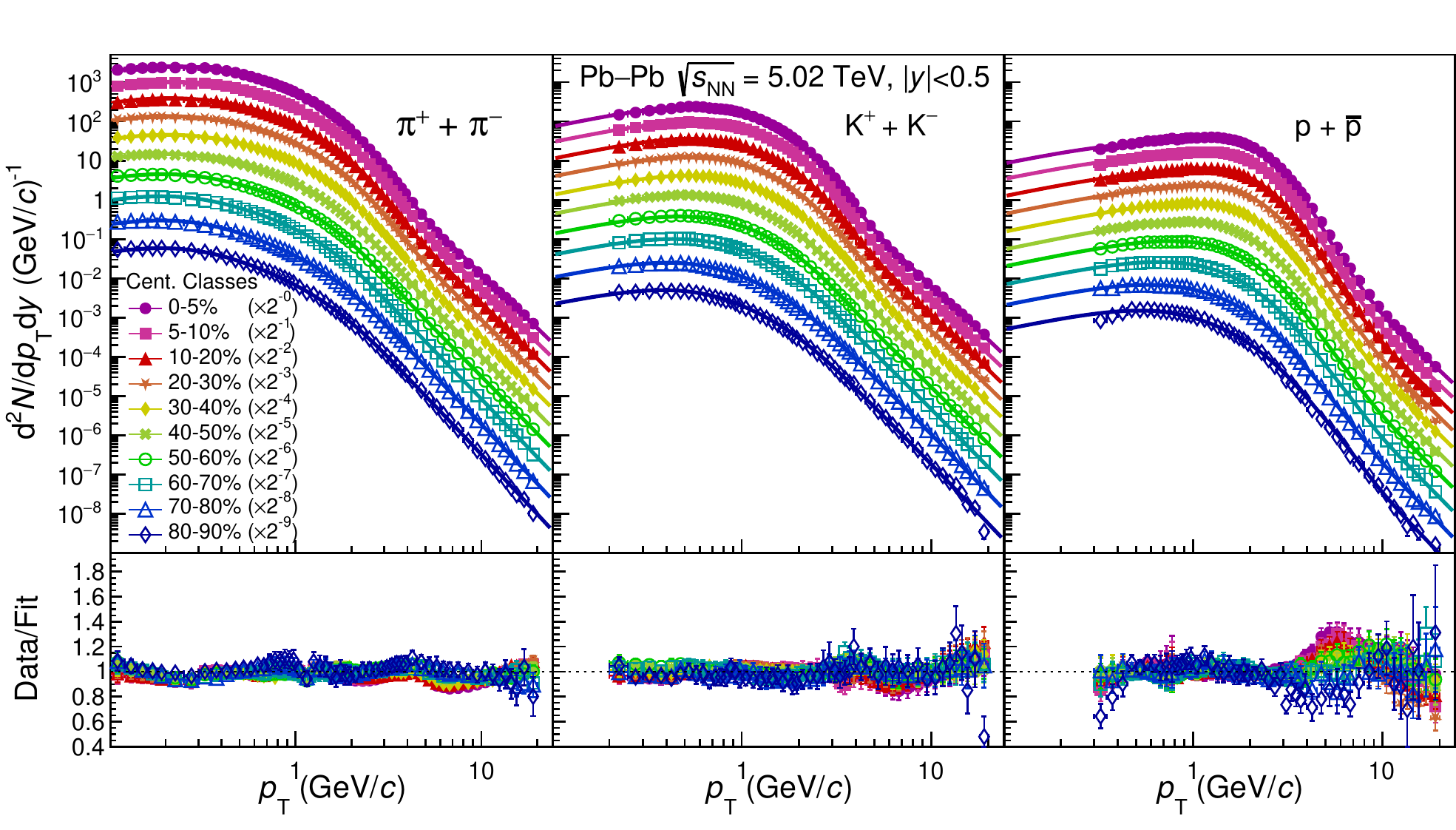}
\vskip -2mm
\caption{Fit of pion, kaon and proton \pt-spectra at $|y| < $ 0.8 (0.5) for different centrality classes in Pb--Pb collisions 
at \enn $=$ 2.76~TeV~\cite{ali18} (5.02~TeV~\cite{ali19}). Numbers in parentheses have the same role as in the Fig. 4.} 
\label{idpPbPbvsM}
\end{figure}
\twocolumngrid  

\noindent
The fit quality almost always is very good.
This is true also for the fits of all other particles.
Thus, our model equally well describes very different shapes of \pt-spectra in pp and Pb--Pb collisions.
Largest deviations between the data and model are mainly 
for the lowest multiplicity (centrality) classes in pp (Pb--Pb) collisions. Note that the experimental definition of such classes has relatively larger uncertainties~\cite{ali2}.

\begin{figure}[ht]
\includegraphics[width=1.03\columnwidth]{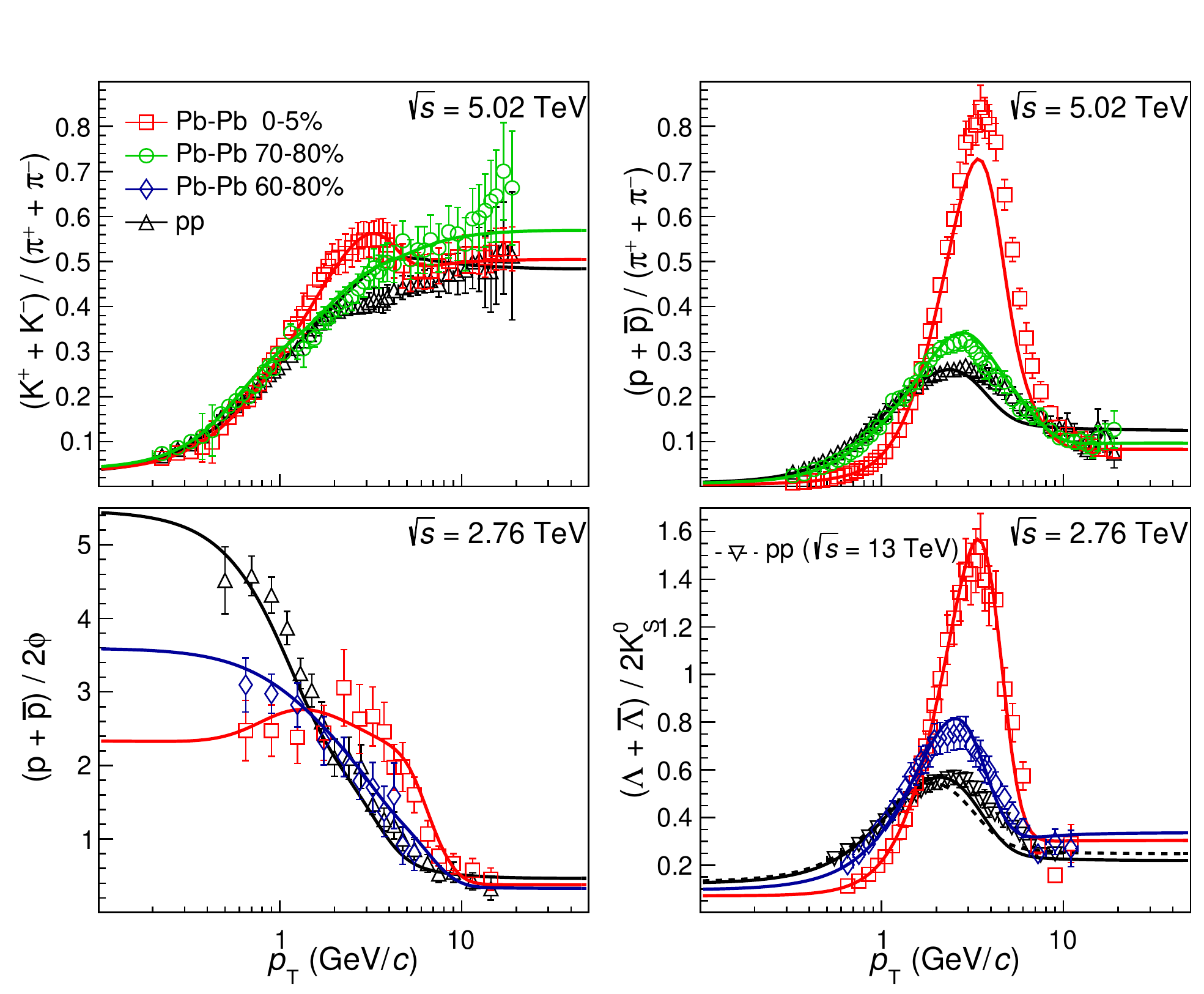}
\caption{Comparison of the model calculations for the ratios 
of different particle \pt-spectra with the ALICE data~\cite{ali10,ali19,ali21,ali24} obtained in Pb--Pb collisions for some centralities at \enn $=$ 2.76 and 5.02~TeV and in inelastic pp collisions at \en $=$ 2.76, 5.02 and 13~TeV.} \label{Rpar}
\end{figure}

An interesting observable, which is very sensitive to the hadron production mechanisms, is the ratio of \pt-spectra of different hadrons produced in the same collisions.
\mbox{Figure 6} displays the comparison of the model
calculations (using the parameter values given in the Sec.~\ref{sec3}) for such ratios with the ALICE data, measured in Pb--Pb collisions 
of different centralities and in inelastic pp collisions.
Since the data for the $(\lmb+\almb)/2\kzero$ ratio are absent in pp collisions at \en $=$ 2.76~TeV, the corresponding data at \en $=$ 13~TeV are shown. A nice agreement is obtained overall.
The peaks in the right panels at \pt $\sim$ 1--7~\gevc are related to the radial flow effects, which are stronger for heavier particles and more central collisions.
Note that the ratios of \pt-spectra are almost independent of \en for LHC energies and reach a plateau at \pt $>$ 20~\gevc.

Another important observable, which measures the suppression of a hadron yield in ion-ion (AB) collisions of some centrality relative to its yield in inelastic pp collisions at the same \en, is the nuclear modification factor

\be  \label{eq:raa}
R_\mathrm{AB} = \frac{\td^2 N_\mathrm{AB} / \td\pt\td y}{\avNcoll \td^2 N_\mathrm{pp} / \td\pt\td y}\, .
\ee
\noindent
Here, \avNcoll~is the number of nucleon-nucleon (NN) binary collisions averaged over the AB events of the given 
centrality and calculated in the Glauber model (see, e.g.,
Refs.~\cite{Loiz,ali28}). 
Factor $R_\mathrm{AB}$ is expected to be equal to unity 
in case of absence of any nuclear effects, when AB collision 
can be considered as a sum of NN collisions.

\begin{figure}[t]
\includegraphics[width=1.05\columnwidth]{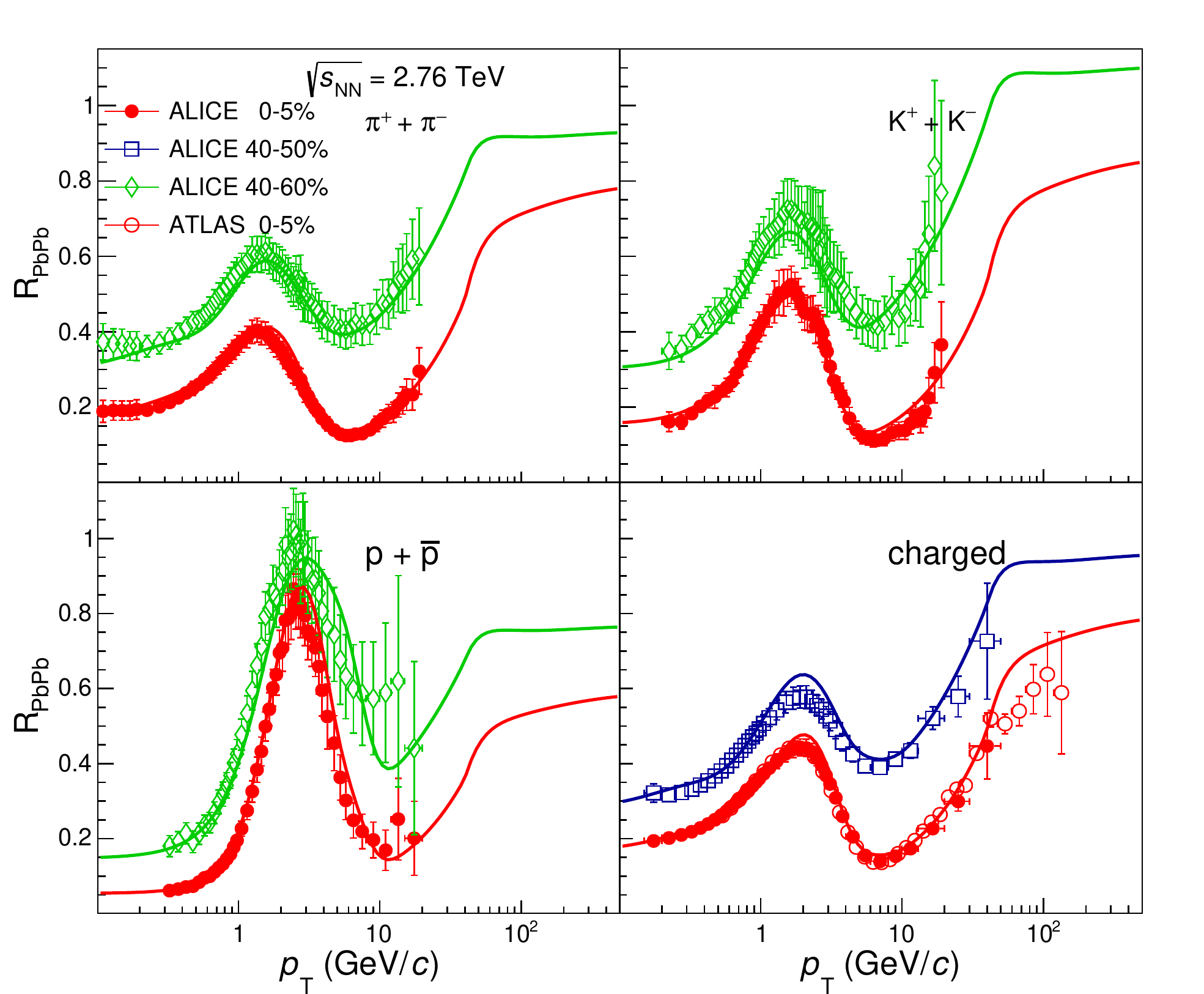}
\includegraphics[width=1.05\columnwidth]{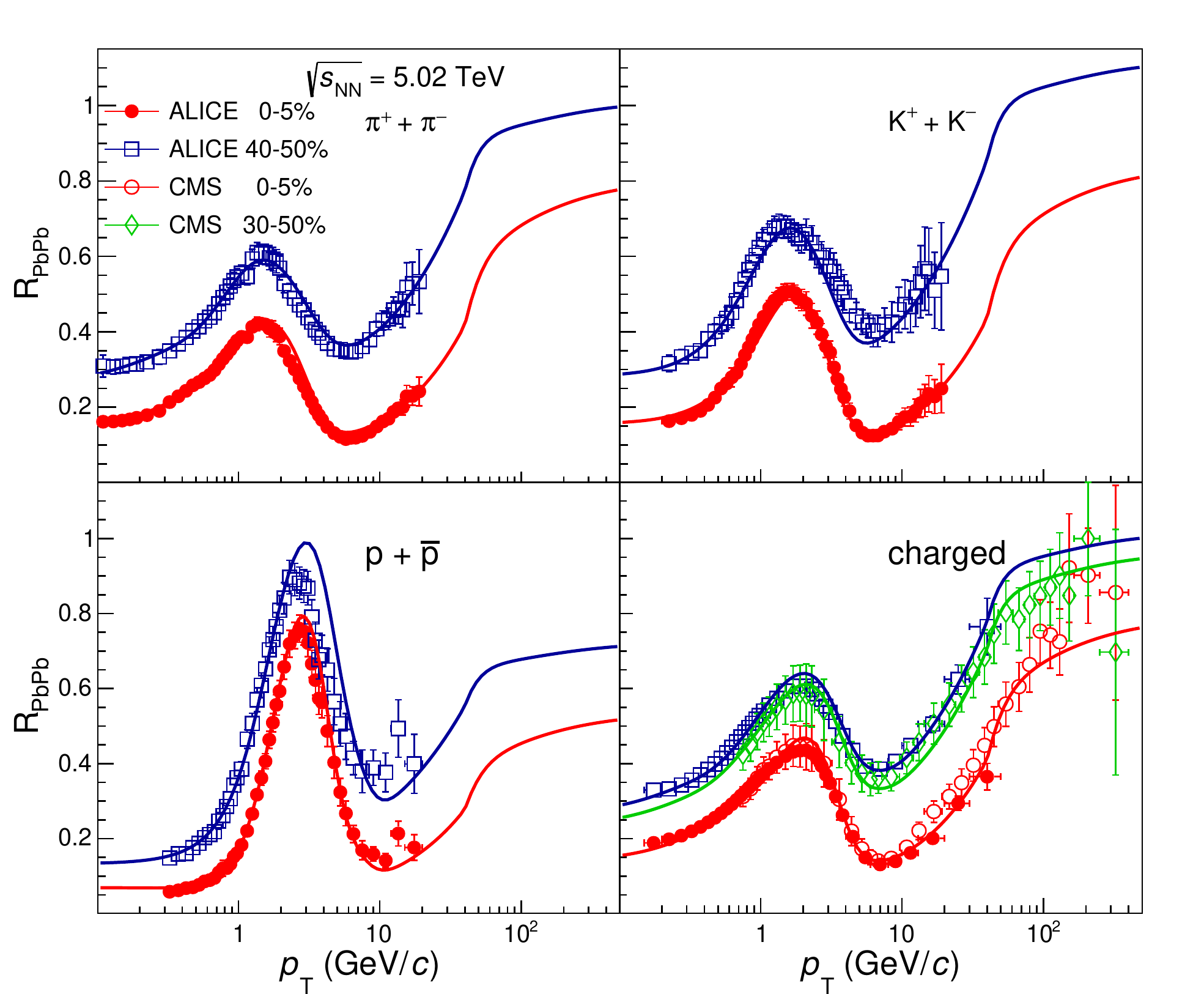}
\caption{Comparison of the model calculations for the nuclear modification factor of different particles with the Pb--Pb data for some centralities at \enn $=$2.76~TeV~\cite{ali18,ali20,atl3} (top panels) and \enn $=$ 5.02~TeV~\cite{ali19,ali20,cms5} (bottom panels).}
\label{RAA}
\end{figure}

Figure~\ref{RAA} shows the comparison of the model calculations for the nuclear modification factors of pions, kaons, protons and unidentified charged-particles with the LHC data for central and peripheral Pb--Pb collisions. A very 
good agreement is achieved. Like in the Fig.~\ref{Rpar}, the peaks at \pt $\sim$ 1--7~\gevc are due to the radial flow effects, which are larger in Pb--Pb than in pp.
The rise in the region \pt $\sim$ 7--40~\gevc,
which is attributed usually to the energy loss of quark-gluon jets in the QGP, is described mainly by the third term of Eq.~(\ref{eq:F}). The further rise at \pt $>$ 40~\gevc and reaching a plateau are described by Eq.~(\ref{eq:hpt}) due to its universal behavior at high \pt. 
For the further checks and improvements of the model it will be important to extend the ALICE measurements for the identified particles \pt-spectra and $R_\mathrm{PbPb}$
into the region \pt $>$ 20~\gevc.

\section{\label{sec7}Conclusion}

Thus, a three component model (BWTPM) is presented in 
Eqs.~(\ref{eq:d2N})--(\ref{eq:chp}), including a standard BWM term, a Tsallis term and a $\pt^2$-dependent power-law term, 
which describes accurately the hadronic \pt-spectra measured at midrapidity in pp and Pb--Pb collisions at the LHC.
It is checked that the modified versions of Eq.~(\ref{eq:F}), with last two terms replaced by two Tsallis forms or two $\pt^2$-dependent power-law forms (as in Ref.~\cite{Byl2}), describe the data worse. 
Another important difference between the model~\cite{Byl2} and BWTPM is that the BWM term dominates at low \pt in the former and mostly at intermediate \pt in the latter. 
This is related to the fact that the model~\cite{Byl2} uses parameters $\bets = 0.5$ and $k=1$, 
while in BWTPM $\bets$ has larger values and $k$ strongly depends on the collision centrality.
In fact, the BWTPM is effectively a two component "soft + hard" model, in view of an auxiliary role of the Tsallis term, which vanishes for heavy particles and presumably is describing mainly the contribution from the resonance decays. 
The model can be further improved by including the resonance decays directly, at least for the BWM component, using the method proposed in Refs.~\cite{Maz1,Maz2}.

It should be noted that BWTPM allows us to make predictions
for new measurements at the LHC.
Since the particle type dependent parameters of Table III are independent of the centrality (multiplicity) classes and \en, 
one can calculate the particle \pt-spectra also for those classes or \en, for which they are not measured yet. 
For instance, spectra of $\Lambda, \Xi^{\pm}$ and $\Omega^{\pm}$ are 
still not measured in Pb--Pb collisions at \enn $=$ 5.02~TeV. Examples of such predictions are shown in Fig.~\ref{compons} for $\Omega^{\pm}$, \Dzero\, and \jpsi in pp at \en $=$ 7~TeV and for $\Lambda, \Xi^{\pm}, \Omega^{\pm}$, \Dzero\, and \jpsi in Pb--Pb at \enn $=$ 5.02~TeV as well as in Fig.~\ref{ch7pi8} for \pipm in pp at \en $=$ 8~TeV.

Application of the BWTPM to describe the LHC data measured
in p--Pb and Xe--Xe collisions will be done elsewhere.
This will allow us to study thoroughly the common trends in the multiplicity dependence of the model parameters across all collision systems from the smallest pp to largest Pb--Pb.


\end{document}